\documentclass{aastex}

\usepackage{graphicx}

\newcommand{\Kepler}{\emph{Kepler\ }}

\usepackage{commath}

\begin{document}

\title{Kepler-108: A Mutually Inclined Giant Planet System}
\shorttitle{Kepler-108: A Mutually Inclined Planet Pair}
\author{Sean M. Mills and Daniel C. Fabrycky}
\email{sean.martin.mills@gmail.com}
\affil{The Department of Astronomy and Astrophysics\\ The University of Chicago\\ 5640 S. Ellis Ave, Chicago, IL 60637, USA}

\begin{abstract}

The vast majority of well studied giant-planet systems, including the Solar System, are nearly coplanar which implies dissipation within a primordial gas disk. However, intrinsic instability may lead to planet-planet scattering, which often produces non-coplanar, eccentric orbits. Planet scattering theories have been developed to explain observed high eccentricity systems and also hot Jupiters; thus far their predictions for mutual inclination ($I$) have barely been tested. Here we characterize a highly mutually-inclined ($I \approx 15-60^\circ$), moderately eccentric ($e \gtrsim 0.1$) giant planet system: Kepler-108. This system consists of two approximately Saturn-mass planets with periods of $\sim$49 and $\sim$190 days around a star with a wide ($\sim300$AU) binary companion in an orbital configuration inconsistent with a purely disk migration origin. 

\end{abstract}

\section{Introduction}

NASA's \emph{Kepler} mission has discovered hundreds of planets and revealed thousands of additional planet candidates likely to be real planets \citep{2013ApJ...766...81F}. The periods, phases, and radii (relative to their host stars) of transiting planets are straightforwardly measured \cite[e.g.][]{2010exop.book...55W}. Transits may only be seen if the orbital plan is nearly edge-on to the observer (i.e., the inclination, $i$, $\approx 90^\circ$). The impact parameter, the distance of closest projected approach between planet and star, can often be determined by the shape of the transit ingress/egress \citep{2003ApJ...585.1038S}\footnote{ However, it is usually difficult to distinguish an inclination of just above 90$^\circ$ just below 90$^\circ$ (both nearly edge-on orbits) with the same impact parameters. In some many body systems it is possible to distinguish these through either dynamical interactions \citep{2013Sci...342..331H} or overlapping mutual transits \citep{2013ApJ...778..185M}.}.

Of the numerous candidates identified, nearly half are found in multiple-transiting planet systems \citep{2014ApJS..210...19B}. The \Kepler data set also has over 100 cases of planets with time-varying orbital periods \citep{2013Mazeh}. These variations are usually attributed to interplanetary gravitational perturbations. These perturbations lead to measurable transit timing variation (TTV) amplitudes for very massive planets, or if planets are close to low-order resonances \citep{2005MNRAS.359..567A}, which many pairs of super-Earths or Neptunes are \citep{2014ApJ...790..146F}. Measurements of TTVs can put tight constraints on planet masses and eccentricities \citep{2011ApJS..197....2F}. 

The absolute nodal angle of bodies on the sky is undetermined by photometry and only relative angles can be constrained due to dynamical interactions or, occasionally, mutual transits. Mutual inclinations can be measured by the change (or lack thereof) in transit duration and depth as a function of time due to orbital plane precession \citep{2002ApJ...564.1019M,2012Sci...337..556C}. Planetary orbits that are highly misaligned will cause rapid orbital plane precession, causing the chord of the transit to move up or down the face of the star. As a result, the chord will lengthen or shrink as it passes through different projected widths of the star, changing the transit duration. Rapid apse precession with very high eccentricities may also cause transit duration and depth changes \citep{2008MNRAS.389..191P}. Combining TTVs, ingress/egress information, and duration/depth changes gives full 3D information on the system, up to a rotation in the plane of the sky.

The vast majority of observed exoplanet systems are statistically consistent with having low ($\lesssim 5^\circ$) mutual inclinations \citep{2014ApJ...790..146F}. Only a few giant planet systems have individually measured mutual inclinations, and these are composed of nearly-coplanar, low eccentricity, often resonant orbits (e.g., GJ 876 \citep{2010ApJ...719..890R}, Kepler-30 \citep{2012Natur.487..449S}, KOI-872 \citep{2012Sci...336.1133N}, Kepler-56 \citep{2013Sci...342..331H}, and Kepler-119 \citep{2015MNRAS.453.2644A}), consistent with a disk migration origin \citep{1980ApJ...241..425G,2002ApJ...567..596L}. The giant planets of our own solar system are also nearly coplanar and thought to have experienced disk migration \citep{2005Natur.435..459T,2007AJ....134.1790M}. Secular chaos may disrupt the architectures of planetary systems after formation and dissipation of the natal disk \cite[e.g.][]{2013arXiv1311.6816D}. This process can lead to highly eccentric and mutually-inclined orbits \cite[e.g.][]{2009Natur.459..817L}, and may be the cause of some of the observed hot Jupiters \citep{2011ApJ...735..109W}. Therefore theory suggests that we may expect to see the signatures of instability and planet-planet scattering in giant planet systems.
However, only two systems are observed to have significant, measured mutual inclinations to date\footnote{There are several known circumbinary systems where the planet is slightly mutually inclined to the binaries and exhibit spectacular precession effects \cite[e.g.][]{2014ApJ...784...14K,2015ApJ...809...26W}, but all seven such currently known systems have low ($\lesssim 5^\circ$) mutual inclinations \citep{2011Sci...333.1602D,2012Natur.481..475W,2012ApJ...758...87O,2012Sci...337.1511O,2013ApJ...768..127S,2014ApJ...784...14K,2015ApJ...809...26W}. Additionally, as such systems are likely to have vastly different histories, here we consider only systems with a single star and multiple planets.}: Kepler-419 b and c are observed to have a marginally detected mutual inclination of $9^{\circ+8}_{-6}$ from TTV and TDV constraints, which is very modest considering the planets' high eccentricities \citep{2014ApJ...791...89D}, and Upsilon Andromeda c and d are reported to have a mutual inclination of $\sim 30^\circ$, based on astrometric measurements using the Hubble Space Telescope fine guidance sensor \citep{2010ApJ...715.1203M}.

Here we present a photodynamic analysis of Kepler-108 (also known as KOI-119 and KIC 9471974) \citep{2014ApJ...784...45R}, a system of two giant planets (Kepler-108b an Kepler-108c, the inner and outer planets respectively) with a large mutual inclination detected through transit duration and depth changes over the \Kepler observing window. In \S \ref{sec:methods}, we describe our methods for identifying the system as one of interest and analysis of its parameters. In \S \ref{sec:results}-\ref{sec:futureobs}, we summarize the results of the analysis, present the system parameters, and discuss what further constraints can be made on the system. We conclude in \S \ref{sec:discussion} with a discussion of the system's dynamics and a general outlook. 

\section{Methods}
\label{sec:methods}

\subsection{Identification of System}

\begin{figure}
\centerline{
\includegraphics[scale=0.8]{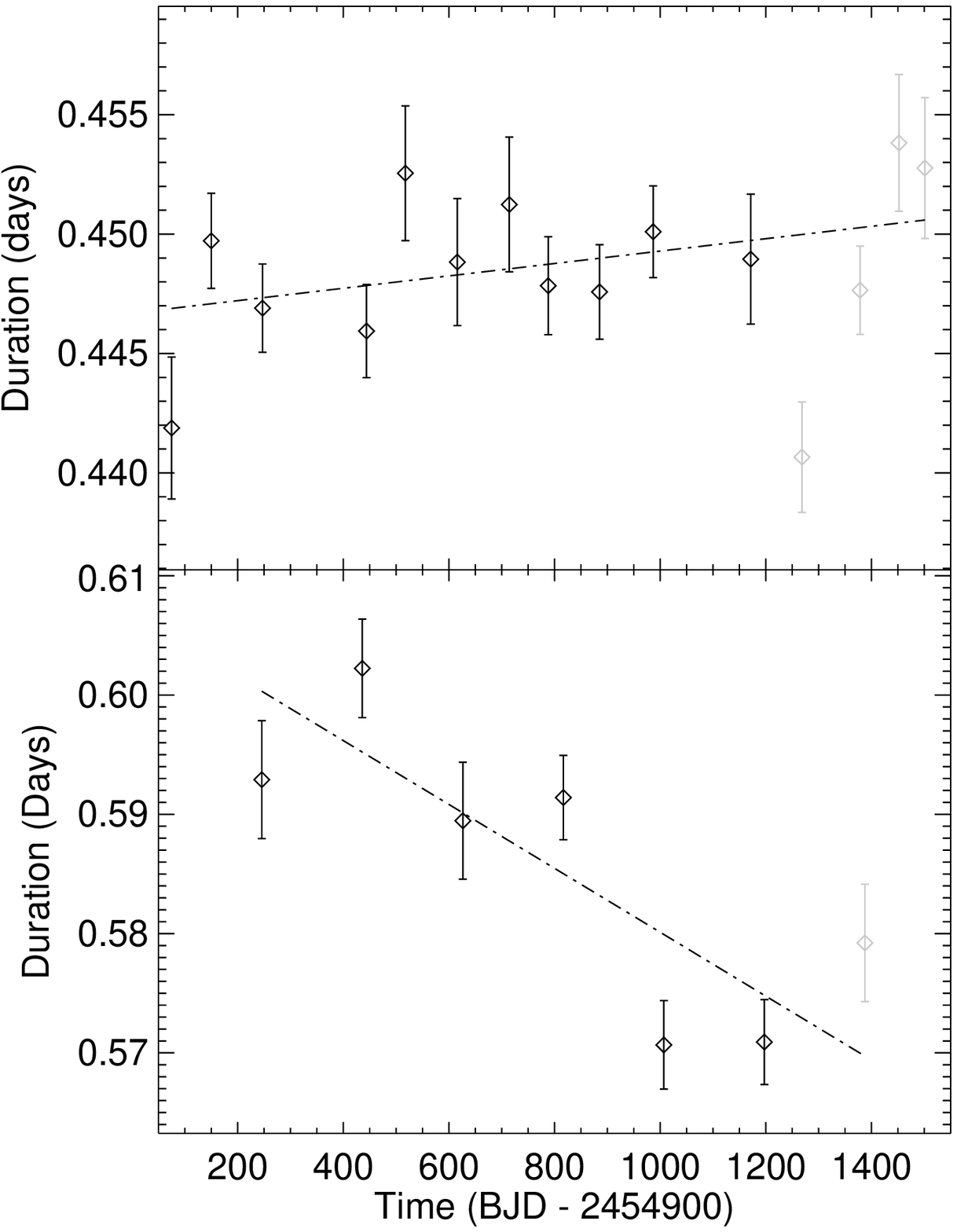}
}
\caption{Transit durations and $1-\sigma$ uncertainties for planet b (top) and planet c (bottom) found by fitting the long cadence data. Data plotted in black represents the 13 quarters used to initially identify the system as one of interest while grey points represent the remainder of the \Kepler data.  A clear trend appears, especially in planet c. }
\label{fig:TDVs}
\end{figure}

\begin{figure}
\centerline{
\includegraphics[scale=0.7]{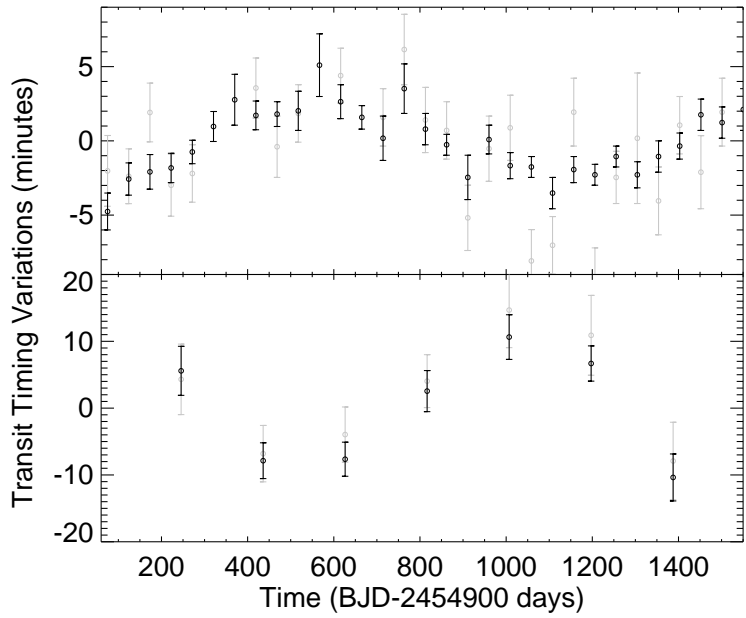}
}
\caption{Individually measured TTVs with $1-\sigma$ uncertainties (gray). Plotted in black are the mean and variance for TTVs measured by taking 100 random draws from the posterior of photodynamical fit as described in \S 3. Therefore the black points combine the \Kepler observational data with a physically possible N-body gravitational model to better constrain the TTVs.  }
\label{fig:TTVs}
\end{figure}

To identify Kepler-108 as a mutually inclined system, we searched the Kepler Object of Interest (KOI) catalog for systems which exhibited possible transit duration variations (TDVs) using the first 13 quarters of data. We began by detrending the simple aperture photometry (SAP) flux data from the \Kepler portal on the Mikulski Archive for Space Telescopes (MAST). For this initial search we used long-cadence (29.4 minute exposure) data. We fit the amplitudes of the first five cotrending basis vectors to determine a baseline and discarded points whose quality flag had a value equal to or greater than 16. In our search, we used the periods given in the KOI catalog and discard any transits within 1.0 days of each other to avoid spurious signals caused by overlapping transits. We fit the individual transits to a time-binned transit function \citep{2002ApJ...580L.171M} and then binned the data into individual \Kepler observing quarters (approximately 3 months). Having thus removed the effects of TTVs from the data, we refit these transits allowing the duration and depth of the transits to vary between quarters. Using an entire quarter of data for each fit allowed lower SNR transits to be fitted and the uncertainties to be small, while still allowing enough distinct data points to see if any duration trends were present. We compute a linear fit to the quarterly best fit durations and compare it to the uncertainty of the duration. Since our data is already subdivided by quarter we can determine by inspection if there are quarterly instrumental issues causing spurious duration changes. A notable concern is that in different observing seasons, a given pixel or group of pixels in the aperture sum might observe a slightly different group of stars causing the transit depth to change as the transit is diluted by background stars. Such effects are readily noticeable as spikes or dips occurring every 4 quarters. Several candidate systems were found in our search, with Kepler-108 being the most convincing, and exhibiting strong TDVs (Fig.~\ref{fig:TDVs}), as well as TTVs (Fig.~\ref{fig:TTVs}).

\subsection{Analysis of Stellar Properties}

An asteroseismology study conducted by \cite{2013ApJ...767..127H} found Kepler-108's mass to be $1.377\pm0.089 M_\odot$. However, Kepler-108 has been the subject of several follow-up studies which have revealed that it is a binary star system.  Adaptive optics (AO) measurements in the i \citep{2014ApJ...791...35L}, J, and K bands \citep{2015ApJ...813..130W} have revealed a companion star 1.05'' from Kepler-108A, which is highly likely to be associated with the system \citep{2015ApJ...813..130W}. The binary nature of the star system is also seen in archival UKIRT images \citep{2007MNRAS.379.1599L}.

To determine which star is the host star, we examine the \Kepler pixel level data and the Data Validation Report (DVR) \citep{2013PASP..125..889B}. \emph{Kepler}'s pixels are approximately 4'' and thus the two sources are not resolved, but we may still determine where within a given pixel a planet's host star lies. Because the field is crowded (including by KIC 9471979, a star within 1 apparent magnitude of the target stars, located approximately 10 arcsec to the south and 3 arcsec to the east), detecting the centroid shift while the planets are in and out of transit is not effective at determining which of the binary stars the planets are transiting because the centroid is affected by these bright, nearby stars \citep{2013PASP..125..889B}. Finding the centroid of the flux difference image (the difference in flux between when the planets are in and out of transit) should reveal the true location of the star being transited, though this method is potentially more uncertain. The DVR indicates that the host star of both planets b and c is approximately 0.5 arcsec East and 0.15 arcsec South of the nominal KIC location at the 4.6-$\sigma$ and 2.2-$\sigma$ levels for each planet respectively. 
Since the KIC location reflects the position of the binary star, its reported position is in between the two observed stars (with the fainter, planet-hosting star lying to the southeast). This is confirmed by centroid fitting of the UKIRT J band images with \texttt{find.pro} and \texttt{starfinder.pro} \citep{2000SPIE.4007..879D} and explains why the centroid offset is only $\sim$0.5 arcsec, rather than the 1.05 arcsec reported by AO imaging \citep{2014ApJ...791...35L,2015ApJ...813..130W}.
We find that stars A and B are located at approximately $(\alpha_A = 294^\circ33'32".085,  \delta_A = 46^\circ03'44.98)$ and $( \alpha_B = 294^\circ33'33".390, \delta_B = 46^\circ03'44".43)$ respectively, whereas the DVR indicates the star hosting planets b and c is located at $(\alpha_b = 294^\circ33'33".334 \pm 0.085, \delta_b = 46^\circ03'44".236 \pm 0.097)$ and $(\alpha_c = 294^\circ33'33".352 \pm 0.155, \delta_c = 46^\circ03'44".267 \pm 0.359)$ respectively. Thus the planets are consistent with each other and Kepler-108B, but not Kepler-108A.
The position angles (PAs) from the KIC location of the host star of planets b and c are $125^\circ \pm 12$ and $111^\circ \pm 37$ respectively, matching the reported PA of the fainter binary companion of $118^\circ$ in both AO images \citep{2014ApJ...791...35L,2015ApJ...813..130W}, and $180^\circ$ from what would be expected from the brighter star.
In summary, this analysis reveals that the position of the planets' host star is consistent with the southeastern star of the binary pair (Kepler-108B), and rules out the brighter star to the northwest (Kepler-108A) at $\gtrsim5\sigma$. 

We use the publicly available Dartmouth stellar isochrone modeling package \texttt{isochrones} \cite[][available at: https://github.com/timothydmorton/isochrones]{2015ascl.soft03010M} to characterize the stars based on the AO flux measurements. We have apparent system magnitudes (taken from https://cfop.ipac.caltech.edu) and the magnitude differences between the two stars in the i \citep{2014ApJ...791...35L}, J, and K \citep{2015ApJ...813..130W} bands. Based on this, we compute the magnitudes of both Kepler-108 stars to use as input parameters to the \texttt{isochrones} package and find that the AO color constraints are consistent within 1-$\sigma$ of the astersoeismology for the brighter star Kepler-108A (the non-planet hosting star), but with a factor of 2.5-6 larger uncertainty in the isochrone method. Lower accuracy is expected from the photometric method because asteroseismology is the most precise method of determining stellar masses and radii developed to date. Nonetheless, the agreement between the different methods of estimations confirms photometry can determine the properties of the planet host star, albeit with large uncertainties. We suggest therefore that our results be interpreted more as broad priors on the scale of the system and stellar density rather than a precise stellar measurement. We summarize our inputs and fitted values in Table~\ref{table:star} and find that the planet-hosting star has $R_\star=0.97_{-0.21}^{+0.56} R_\odot$ and $M_\star = 0.96_{-0.16}^{+0.29} M_\odot$.

\subsection{Photodynamic Analysis}

We followed up our initial analysis of Kepler-108 by applying a photodynamic model. The model integrates the 3-body Newtonian equations of motions for the central star and two planets, including the light travel time effect. When the planets pass between the star and the line of sight, a synthetic light curve is generated \citep{2012MNRAS.420.1630P}, which can then be compared to the data. For the photodynamics, we took advantage of the short cadence (58.8 second exposure) data available in \Kepler Quarters 5-8 and 12. Cotrending basis vectors are not available for short-cadence data. To detrend this data, first we masked out the expected transit times (as done for the long cadence data) and then fit a cubic polynomial model with a 1-day width centered within half an hour of each data-point, to determine its baseline. We divide the flux by this baseline. We continued using long cadence data where short cadence was not available (\Kepler Quarters 1-3, 9-11, and 13-17). We used the photodynamic model to produce theoretical normalized flux values at the timestamp of each short cadence data point. For long cadence data, we computed the flux value at 15 equally-spaced points in time over a cadence's integration and averaged them together to produce the theoretical result. A small amount of correlated noise (fractional variations $\lesssim 10^{-4}$, with a peak in the Lomb-Scargle periodgram near 45 minutes) was still present in the data, likely due to the stellar variability which allowed for asteroseismology measurements. Our detrending algorithm did not address this short time-scale noise to avoid distorting the transit shapes. Therefore we multiplied the quoted data uncertainties by a multiplicative factor of 1.075 such that the reduced $\chi^2$ of our best-fitting model was 1.0.\footnote{A more careful treatment could, in principle, be done using Gaussian process noise modeling \cite[e.g.][]{2014arXiv1403.6015A}, however this was computationally untenable for our study since we have $\sim 6\times 10^5$ data points for a model that needed to be run $> 10^9$ times for all of the different DEMCMCs. } One of the planets, Kepler-108b, had only partial transits observed at $\sim$BJD 2455959, 2456106, and 2456204 due to pauses in data collection. Our detrending algorithm performs poorly on cases where there is not a baseline on both sides of the transit. Therefore the data within 1 day of these transits was removed from the fits to avoid incorrectly influencing the fit by changing the measured depths. Since the mid-time and duration measurements of these transits is highly uncertain due to having only either the ingress or egress, retaining them would add minimal information to our fits. In total we were left with 612057 photometric data points.

The parameters for each planet in the differential evolution Markov chain Monte Carlo \cite[DEMCMC, ][]{TerBraak2005} fit are $\{P, T_0, e\cos(\omega), e\sin(\omega), i, \Omega, R_p/R_\star, M_p/M_\star\}$, where $P$ is the period, $T_0$ is the mid-transit time, $e$ is eccentricity, $\omega$ is the argument of periapse, $i$ is inclination, $\Omega$ is nodal angle, and $R$ and $M$ are radius and mass respectively (with subscripts $p = b,c$ for the planets and $\star$ for the star). The star had five additional parameters: $\{M_\star, R_\star, c_1, c_2, dilute\}$, where $c_i$ are the two quadratic limb-darkening coefficients and $dilute$ is the amount of dilution from other nearby sources. 

The relative flux from each star in the \Kepler bandpass is also uncertain, so we allow the fractional amount of flux from the non-host star to vary as a free parameter. We ran photodynamic fits assuming Kepler-108B is the host with the the stellar mass fixed at the value found as described in \S2.2 because photometry alone can only determine the stellar density. Our DEMCMC fits also used the measured stellar radius and uncertainty ($R_\star =  0.97_{-0.21}^{+0.56} R_\odot$) as a data point along with the \Kepler photometry. Since $dilute$ is highly degenerate with the size of the planets ($R_p/R_\star$), our planetary radii are significantly more uncertain than previously reported values, which did not take into account contamination from another blended source. The shape of the transit does offer some constraints on the dilution so it is not a completely degenerate parameter. We do not include the companion star in our photodynamic model directly because its great distance \cite[at minimum, the measured sky-projected distance of 327 AU,][]{2015ApJ...813..130W} prevents it from detectably influencing the Kepler-108 planets over the \Kepler observing window. We discuss potential long-term effects in \S6. 
Although it is disfavored, we also ran a second, nearly-identical set of DEMCMCs assuming that the asteroseismologically measured star is the host star, and thus fixed stellar mass at $1.377 M_\odot$ and used the constraint $R_\star=2.192 \pm 0.121 R_\odot$ \citep{2013ApJ...767..127H}. We include these posteriors in the appendix.

To test whether our detection of changing transit durations and depths (and therefore mutual inclination) was robust, we ran 3 different DEMCMCs for each host star: ($\mathcal{P}1$) allowing the inclinations and relative nodal angle of the planets to vary independently; ($\mathcal{P}2$)  allowing only the planets inclinations to vary independently and fixing both planets to a nodal angle of $\Omega=0$; and ($\mathcal{P}3$) forcing the planets to be coplanar, i.e. fixing $\Omega=0$ and $i_b=i_c$, but allowing the value of the inclination to vary.

Requiring strict coplanarity ($\mathcal{P}3$) results in a far worse fit to the data ($\Delta \chi^2 > 150$) than the other two models ($\mathcal{P}1$ and $\mathcal{P}2$) regardless of host star, because each planet must have the same inclination. As a result, the impact parameter of both planets is determined by a single inclination, and the transit shapes and durations of both planets can not be fit well compared to the case where two different inclination values are allowed. We no longer discuss $\mathcal{P}3$ as a viable candidate model since with only one additional free parameter we vastly improve the fit and provide a more realistic model.

The $\mathcal{P}2$ ($\Omega_{b,c}=0$) DEMCMC was initialized with the periods of the two planets as reported in the \Kepler catalog \citep{2013Batalha}, and at a variety of eccentricities below $0.1$ for both planets. The DEMCMC chains slowly explored increasingly higher eccentricities, with the chains preferring for the inner planet (b) to have $e > 0.7$. Once the DEMCMC chains found this high eccentricity space, they did not travel back to lower eccentricities because high eccentricity allowed a better fit to the data. Restarting the DEMCMC from a variety of solutions with planet b having $e\sim0.75$ and $\omega \sim 150^\circ$ (near the best fit found previously), resulted in none of the chains seeking lower eccentricity regions. Therefore we conclude that for the nearly coplanar case ($\mathcal{P}2$), high eccentricity solutions are robustly preferred. We ran a parallel 46-chain DEMCMC until the the parameters appeared stationary, the chains were well mixed ($> 45$ autocorrelation time scales for every parameter for each chain on average), and their Rubin-Gellman $\hat{R}_{interval}$ statistic \citep{Brooks98} was below 1.2 for every parameter. We recorded the parameters for each chain every 1000 generations for $5\times10^6$ generations to reduce correlation and required disk space, and threw out the first $5\times10^5$ generations of all chains as a burn-in. We thus obtained $2\times10^5$ samples of the posterior for each DEMCMC, of which at least $45\times46\approx2000$ are completely independent.

\begin{figure}
\hspace{0.5cm}
\centerline{
\includegraphics[scale=0.15]{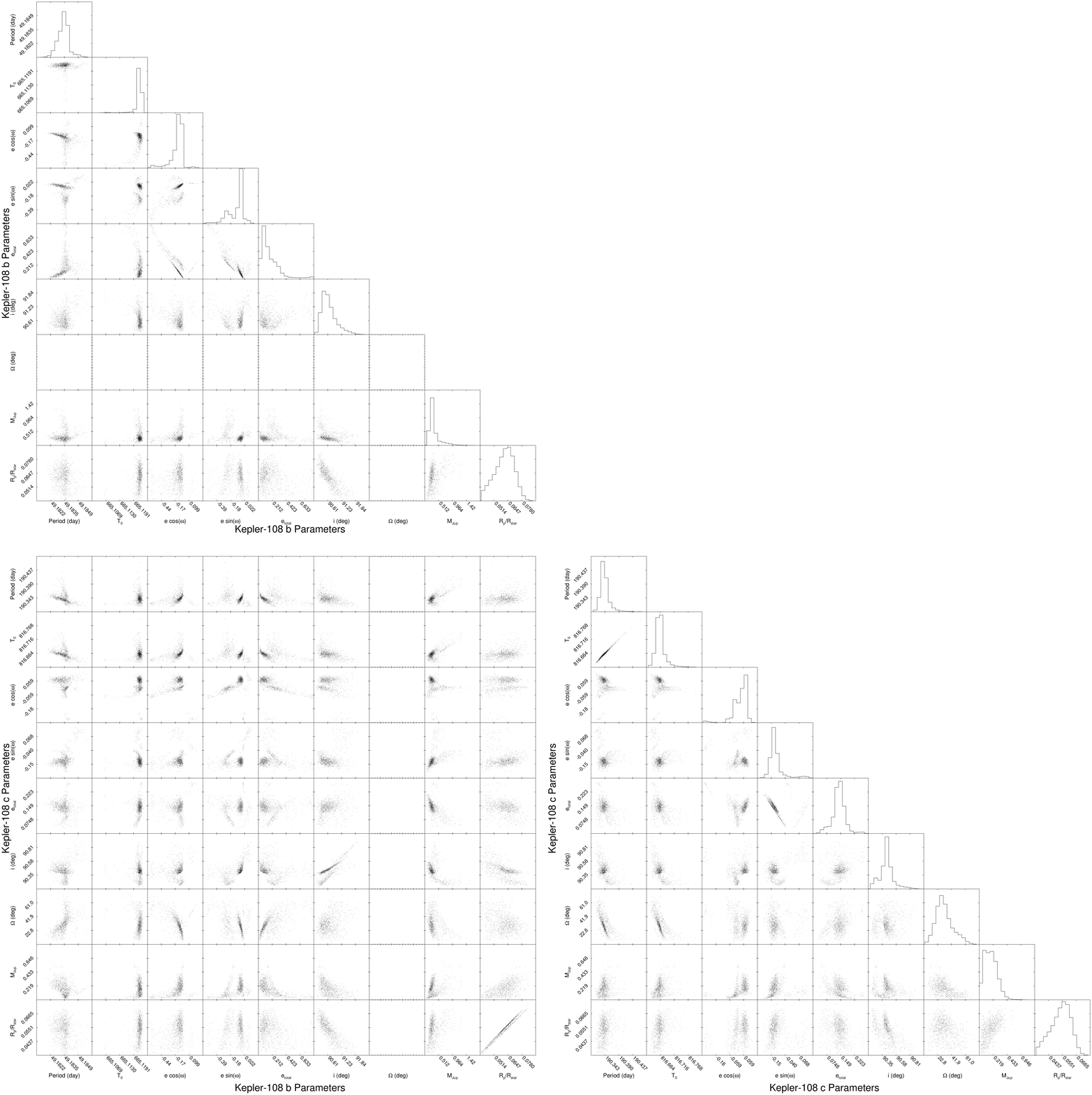}
}
\caption{
Correlations among all planetary parameters in $\mathcal{P}1$, the mutually inclined model. Where correlations would be between a parameter and itself, instead a histogram of the distribution of that parameter is shown. 
 }
\label{fig:planetcorr}
\end{figure}

\begin{figure}
\centerline{
\includegraphics[scale=0.25]{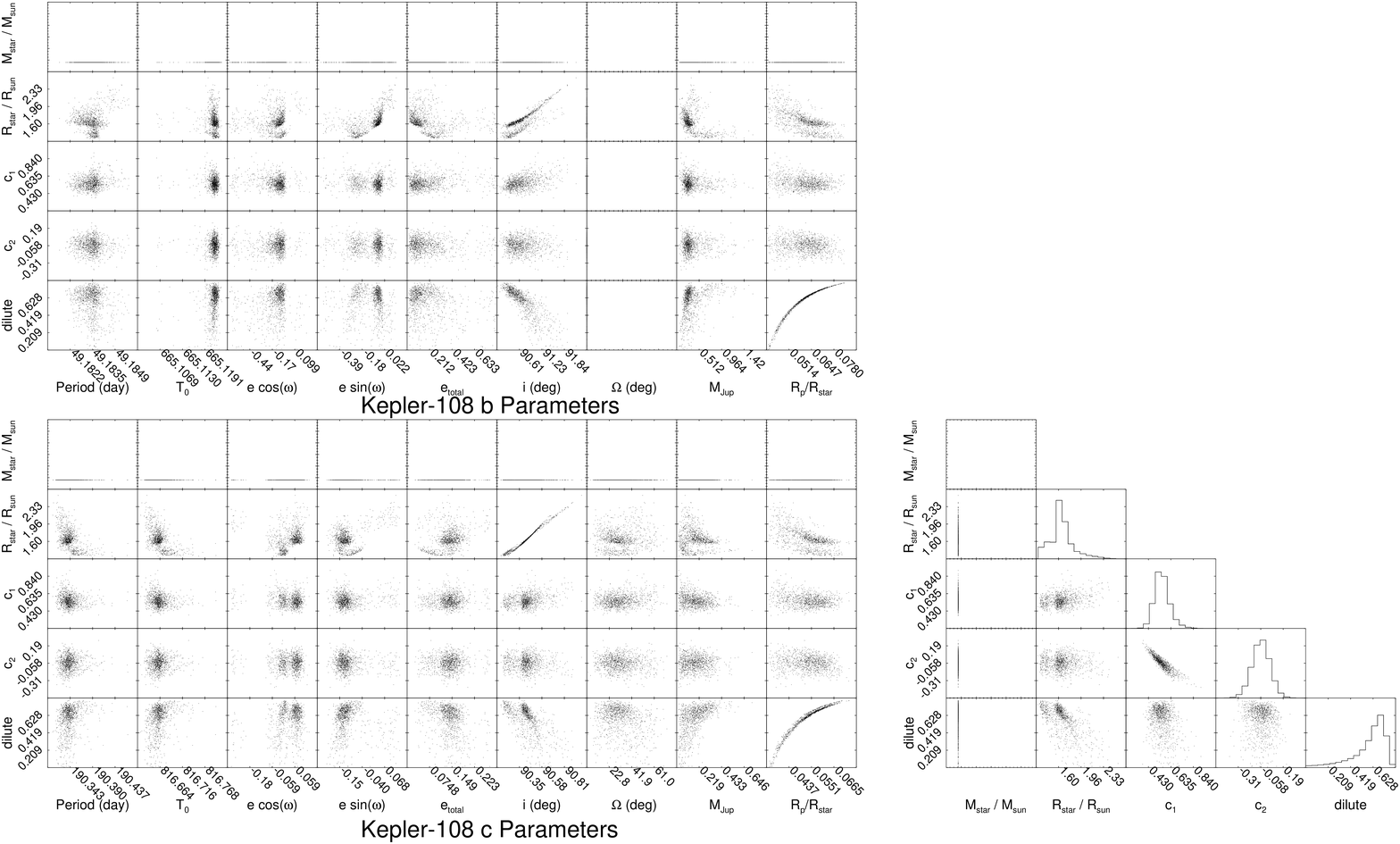}
}
\caption{
All correlations involving stellar parameters in $\mathcal{P}1$, the mutually inclined model. Where correlations would be between a parameter and itself, instead a histogram of the distribution of that parameter is shown. 
 }
\label{fig:starcorr}
\end{figure}

The $\mathcal{P}1$ DEMCMC was initialized similarly to $\mathcal{P}2$, with all eccentricities below $0.1$. This DEMCMC also explored higher eccentricities for planet b, but rather than remain high as in $\mathcal{P}2$, eccentricities continually varied between high and low values ($0.0 \lesssim e \lesssim 0.7$). Concerned that this DEMCMC was not finding the same very high eccentricity parameter space found by $\mathcal{P}2$, we ran $\mathcal{P}1$ again but starting from solutions drawn from $\mathcal{P}2$. All chains in this case quickly found lower eccentricity solutions and did not return to the very high eccentricity starting conditions. We ran this ($\mathcal{P}1$) DEMCMC for $3.2\times10^7$ generations, with a $5\times10^6$ generation burn-in. This DEMCMC was run much longer because the wide range of acceptable eccentricities caused slower convergence. When the DEMCMC was stopped, each parameter had experienced $> 20$ autocorrelation time scales (at least 900 independent points) and had $\hat{R}_{interval}$ statistic below 1.2. These values are still acceptable for convergence and continuing running was not computationally feasible. The complex nature of the parameter space (see Figs.~\ref{fig:planetcorr} and~\ref{fig:starcorr}) slows down the convergence significantly, particularly at high eccentricity.
DEMCMC runs with Kepler-108A as host have similar statistics.

Concerned that we could potentially miss additional minima distant from our initialization on the $\chi^2$ surface, we also ran a 4-temperature parallel-tempered DEMCMC with both the $\mathcal{P}1$ and $\mathcal{P}2$ constraints. This approach allows the highest temperature chains to traverse local maxima and explore a much broader range of parameter space. The high temperature chains may then swap with low temperature chains once near a new minima and allow for a more complete exploration of parameter space \cite[see, e.g.,][for more details]{2005PCCP....7.3910E}. We find very similar posteriors and no additional minima which would affect our fits with this method.

\section{Photodynamic Results}
\label{sec:results}

\begin{figure}
\centerline{
\includegraphics[scale=0.7]{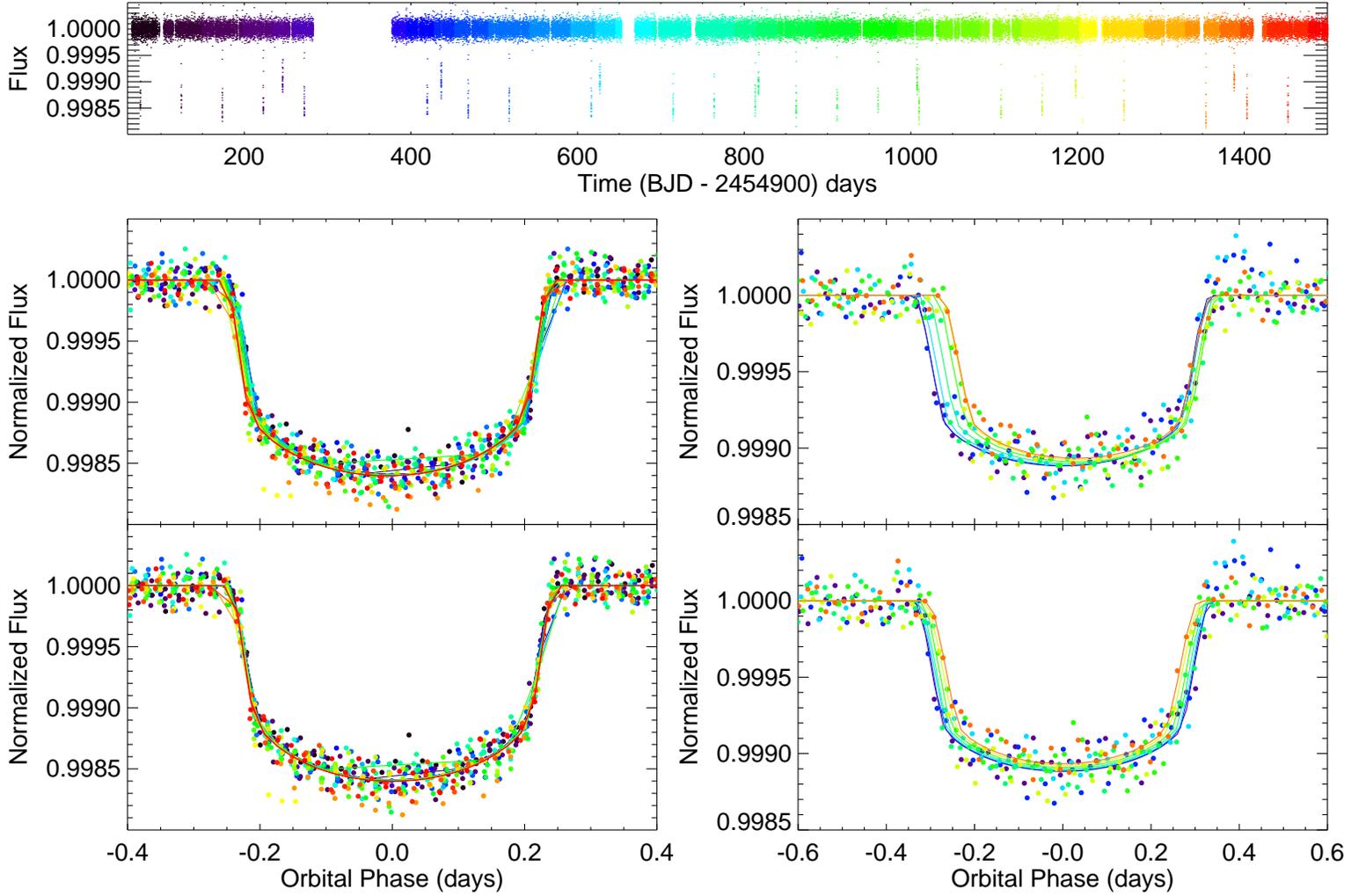}
}
\caption{{\em Top:} Detrended flux over the Kepler observing window. The two planet transits appear clearly as periodic dips of different depth. Color is changed incrementally from violet to red such that each transit has a distinct color in the data. To reduce visual scatter, only long-cadence data is displayed although short-cadence data was used where available in the fitting procedure.  {\em Bottom:} Left and Right columns are planets b and c respectively. The top panel shows the data (dots) and photodynamic best fit model (line) phase-folded with a constant period (the best fit at $T_{epoch}=640.0$ (BJD-2454900), see Table~\ref{table:bestfit}). Bottom panels show the transits phase folded with the TTVs removed. This allows clear identification of the change in planet c's duration and depth with time (as indicated by color in the top panel). Model points are produced only where real data points are found and are connected by straight lines resulting in the apparent sharp corners on some of the transits.}
\label{fig:stacktrans}
\end{figure}

The data generally allow for two classes of solutions which cause the observed duration and depth changes in the transits (see Fig.~\ref{fig:stacktrans}). The first case, explored by $\mathcal{P}1$, we will describe as the low eccentricity, high mutual inclination case.
The second case, explored by $\mathcal{P}2$, refers to the nearly coplanar, highly eccentric case. In this case, the mutual inclination between the planets is $\lesssim 1^\circ$. The very large ($\sim 0.75$) eccentricity of the inner planet along with its increased mass causes faster precession of the node (and apse) of the outer planet. Along with the larger eccentricity of the outer planet, this results in similar transit duration and depth changes. DEMCMC posterior median values and $1-$ and $2-\sigma$ uncertainties at $T_{epoch}=640.0$ (BJD-2454900) for the mutually-inclined ($\mathcal{P}1$) and nearly-coplanar ($\mathcal{P}2$) models are given in Table~\ref{table:allparams}. Note that the distributions in many parameters are not Gaussian and the $2$-$\sigma$ interval is generally not twice as wide as the $1$-$\sigma$ interval. The distributions and correlations between parameters for $\mathcal{P}1$ are shown in Figs.~\ref{fig:planetcorr} and~\ref{fig:starcorr}. Correlations in other fits are similar. Confidence intervals higher than $2$-$\sigma$ are not given as the number of independent parameter values mean that there are relatively large fractional uncertainties on the confidence intervals of higher $\sigma$; however, the two sets of confidence intervals given are sufficient to understand the posteriors. Best fit solutions found by DEMCMC under $\mathcal{P}1$ (first) and $\mathcal{P}2$ (second) constraints assuming Kepler-108B as the host star at $T_{epoch}=640.0$ (BJD-2454900) with $\chi^2 = 609823$ and $609850$ respectively are given in Table ~\ref{table:bestfit}. 

Massive, high-eccentricity planets will strongly perturb each other; however, in a random sample of 100 draws from each posterior distribution, 96\% were stable for $> 10^7$ years in $\mathcal{P}1$ and 100\% were stable over the same time period in $\mathcal{P}2$, so stability alone can not easily rule out either regime. The unstable draws from $\mathcal{P}1$ all had one or both planets with higher eccentricity than the 1-$\sigma$ confidence interval where the majority of the posterior lies. 

The fixed nodal angle solution ($\mathcal{P}2$) around Kepler-108B had a best fit $\chi^2=609850$ for 612058 data points, and the system allowing non-zero mutual nodal angles ($\mathcal{P}1$) had $\chi^2=609823$. Since only 1 additional free parameter is added from the $\mathcal{P}2$ to $\mathcal{P}1$ models, we would expect an improvement of $\chi^2$ of order unity if both models fit the described the data well, i.e. if it were true that large mutual inclination was not required to fit the data effectively \citep{1974ITAC...19..716A}. The large difference in $\chi^2$ suggests that the fit allowing large mutual inclinations is superior to the others by $> 4 \sigma$ as follows. 

Rigorously, we can define the F-ratio as the improvement in $\chi^2$ normalized by the number of new free parameters:
\begin{equation}
\Delta \chi^2 / \Delta DOF = (609850 - 609823) / (18 - 17) = 27
\end{equation}
to the final reduced $\chi^2$:
\begin{equation}
\chi_f^2/\nu_f = 609823/612058 = 0.9963
\end{equation}
The F-test gives the probability (p-value) that the F-ratio is as high as observed by chance. In our case the p-value is $2\times10^{-7}$, so we may reject that the planets have the same nodal angle on the sky. We note that the $\chi^2$ being slightly below $1.0$ suggests we have overestimated our uncertainties and therefore only strengthens our reasoning. 

To compare the entire distribution of parameters found by MCMC rather than just the best-fit solution, we computed the Bayes Factor, $K$, using Newton and Raftery's $p_4$ estimator \citep{newton1994approximate} and found the odds ratio to be $>10^{10}$ in favor of $\mathcal{P}1$, i.e. large mutual inclination is strongly favored \citep{kass1995bayes}.

Lastly, a physical argument can be made in support of $\mathcal{P}1$. The radii of the two planets of Kepler-108 differ by only $\sim$20\% (in all scenarios). In the $\mathcal{P}1$ model, the planet masses differ by a significant, but reasonable, factor of $\sim$2.7. In the $\mathcal{P}2$ model, the masses must differ by a factor $\sim$70, implying that planet c, with a radius $R_c \approx 0.7 R_{Jupiter}$ has a mass of only $M_c \approx 0.017 M_{Jupiter}$, which implies a lower density than all but the most extreme sub-Neptune planets \citep{2014ApJ...783...53M}. 

Here we have compared the coplanar models only for the case where Kepler-108B is the planet host. However, we perform an identical analysis for the case with Kepler-108A as the planets' host, and this analysis also favors the mutually inclined case to a similar significance. Thus, even if there is some doubt regarding which star the planets orbit, we may say unambiguously that the planets are mutually inclined.

\section{Observation Statistics}

\begin{figure}
\centerline{
\includegraphics[scale=1.0]{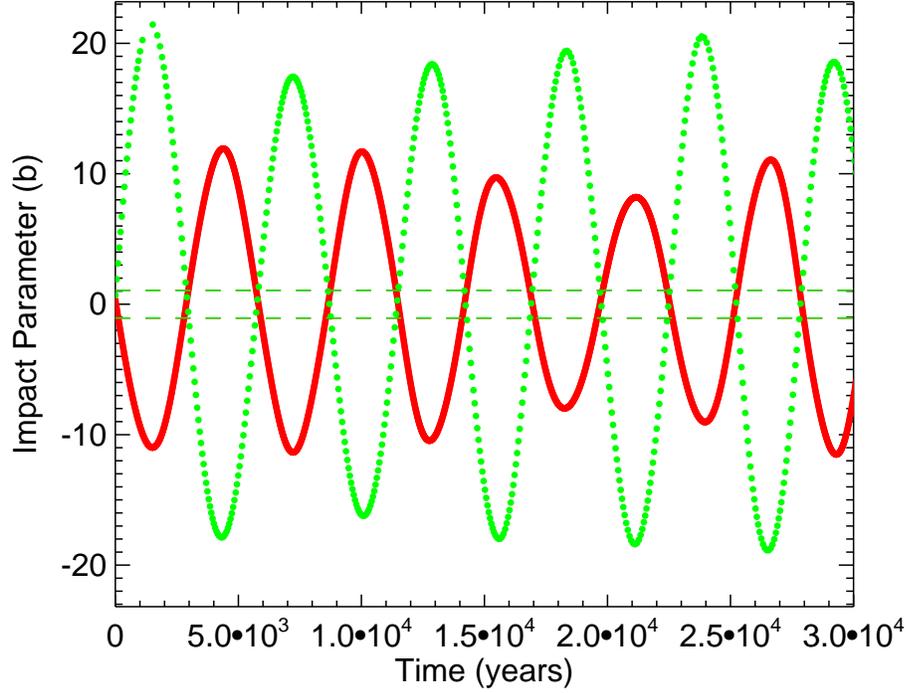}
}
\caption{Evolution of the impact parameter ($b$) of both planets of Kepler-108 over $3\times10^4$ years. While $b$ is usually reported as a positive definite quantity, we have assigned a negative value for $b$ whenever the position of the planet at minimum $b$ for a given transit is below the center of the star (negative $y$ value). This allows us to visualize the planet moving up and down, on and off the star. Dashed lines show the maximum $b$ where the planet will transit ($b_{max} = (R_\star+R_i) / R_\star$, $i=b,c$). 
This data is taken from a portion of the $10^5$ year run of the best-fit solution (see Table~\ref{table:bestfit}).  
The asymmetry with respect to $b=0$ is due to the invariant plane being inclined to the observer. }
\label{fig:bchange}
\end{figure}

Since the planets in this system are precessing due to the high mutual inclination between them, the planets will eventually change their orientation so dramatically that they no longer transit. This has been observed in circumbinary systems \citep{2014ApJ...784...14K,2015ApJ...809...26W}, but never in a single-star planetary system. Few other known extrasolar systems are likely to exhibit large inclination variations due to self-excitation \citep{2016MNRAS.455.2980B}.
To investigate the timescale of the precession in this system, we integrate the best-fit solution forward for $10^5$ years (Table \ref{table:bestfit}). We find that both planets periodically precess on and off the star (see Fig.~\ref{fig:bchange}). 
From our viewing perspective, this system has 2 planets transiting 3\% of the time, 1 planet transiting 4\% of the time, and no observable transits 93\% of the time.
The precession timescale, $P_{prec}$, is found numerically to be on average $\sim5700$ years, a bit longer than an analytic prediction $\sim4400$ years using the frequency for $\Omega$ and $i$ oscillations found by applying the Laplace-Lagrange secular solution to first order in planet mass and second order in inclination \cite[e.g.][]{murray1999solar}. 

In order to better understand the statistics of observing systems like Kepler-108, we explore the the likelihood that this system is observed as two transiting planets experiencing TDVs from any orientation. We track the position in 3-dimensional space of both planets in our best fit model every minute for one complete orbit of the outer planet at the beginning and end of the \Kepler observing window. That is, we produce two $\vec{x}(t)$ functions for each planet ($\vec{x}_{b,1}(t)$, $\vec{x}_{b,2}(t)$, $\vec{x}_{c,1}(t)$, and $\vec{x}_{c,2}(t)$) each $190$ days long and $\Delta t \sim$$1300$ days apart. We then randomly draw 10,000 different observing orientations and compute the impact parameter ($b_{j,k}$, $j=b,c$, $k=1,2$) for each planet ($b$ and $c$) in both windows ($1$ and $2$) from each orientation. We compute the implied duration ($D_{j,k}$) of the transit corresponding to each $b_{j,k}$ using \citep{2010exop.book...55W}:
\begin{equation}
D_{j,k} = \frac{P_j}{\pi} \frac{\sqrt{1-e_j^2}}{1-e_j \sin(\omega)} \sin^{-1}\bigg(\frac{\sqrt{(R_\star+R_j)^2-b_{j,k}^2}}{a_j \sin(i)}\bigg) 
\end{equation}
where the orbital elements come from the instantaneous position and velocity of the planets at the time of minimum $b$. This is a good approximation for the true duration. The change in duration over the observing window $(\Delta D)_j$ is given by $D_{j,2} - D_{j,1}$. We establish as a detectability threshold $(\Delta D)_j = 30$ minutes (the approximate limit of a confident detection of duration change in Kepler-108) and compute the fraction of observation angles for which $(\Delta D)_j$ exceeds the threshold. We use the same threshold for both planets since they are approximately equal in radius, i.e. transit signal. 

The results of this analysis are summarized in Table~\ref{table:obsfrac} which lists the fraction (and uncertainty) of randomly chosen viewing angles for which the Kepler-108 system would be observable as a 2-planet system, 1-planet system, and a 0-planet system by the \Kepler mission. Because the planets are highly mutually inclined, seeing a single planet transit does not guarantee that the second will be visible. This is seen in the simulations as the 2-planet observations are much fewer in number than the 1-planet observations, which are dominated by the interior planet due to its closer orbit to the star. The second column shows the fraction of viewing angles for which Kepler-108 would appear to have duration variations in either planet of greater than 30 minutes (a rough limit on a confident detection of the duration change in Kepler-108). Approximately half of the cases where 2 planets are visible show measurable duration drift, however in the case where only 1 planet is visible, measuring a duration drift will happen only $\sim$8\% of the time.

It is clear from these statistics that our current viewing geometry is unusual. Since we have observed Kepler-108 as a 2-planet system exhibiting TDVs, it is probable that we have also observed similar systems in different viewing configurations. In other words, it is likely that some observed single Jupiter systems may actually be members of mutually inclined multi-Jupiter systems. Thus, unless we are very unlucky, we expect that a close analysis of many systems with a single transiting Jupiter will reveal duration and depth changes in a few systems due to a non-transiting, mutually-inclined companion. However, the measurement of a single planet's duration change gives very degenerate information about the perturbing planet's parameters, and it is more challenging to rule out systematics without a well-defined perturbing planet.

\section{Future Observations}

\label{sec:futureobs}

To assist potential future follow-up measurements, we predict TTVS and 1-$\sigma$ uncertainties based on 100 random draws from the $\mathcal{P}1$ posterior up to 10 years after the end of \Kepler data collection (Table~\ref{table:ttvs}).

\subsection{Spin-Orbit Alignment}

There is limited observable star spot activity on Kepler-108 in the \Kepler data due to low SNR. Thus identifying the alignment of the stellar spin with the planets' orbits was not possible using star spot crossings \citep{2011ApJ...740L..10N}. Previous spectroscopic measurements of Kepler-108 gave $v \sin(i)= 5.3 \pm 0.6$ km/s where $i$ is the inclination of the stellar spin axis to the line of sight and $v$ is the star's rotational velocity \citep{2013ApJ...767..127H}. For a star of radius $R_\star = 2.192 R_\odot$, this suggests a maximum rotation period ($i=90^\circ$) of $\sim$$22.4$ days, but provides little information regarding the star's inclination relative to the observer. More importantly, it is not clear for which of the two stars in the binary this measurement is relevant. 

The sky-projected angle between the stellar spin axis and the planets' orbit normals can be measured spectroscopically by identifying the change in apparent radial velocity of the stars as the planet crosses \cite[known as the Rossiter-McLaughlin effect, see, e.g.,][]{2007ApJ...655..550G}. The expected Rossiter-McLaughlin amplitude for the observed spin is $K_R = 6.9$ m/s, which is potentially observable \cite[see e.g.][]{2015arXiv150301770P}, though the transits are quite lengthy. These planets likely went through some chaotic destabilization event to get into mutually inclined orbits from their presumably coplanar, protoplanetary disk formation configuration. We therefore predict that the planets could be highly misaligned with the star's spin-axis\footnote{This would not be particularly abnormal, even without the large mutual inclination, because of the high mass of the star. Kepler-108 is an evolved F-type star, a spectral type which commonly exhibits misalignment between planet orbits and stellar-spin \citep{2010ApJ...718L.145W,2015ApJ...801....3M}.}.

\subsection{Radial Velocity Constraints}
\begin{figure}
\centerline{
\includegraphics[scale=0.5]{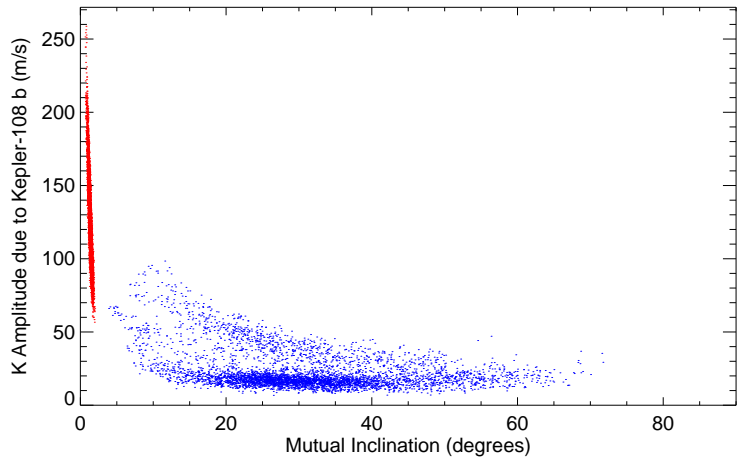}
}
\caption{
Theoretical K amplitude of the inner planet ($P \approx49.2 $ d) as a function of mutual inclination of the two planets for both $\mathcal{P}1$ (blue) and $\mathcal{P}2$ (red). Plotted are 10,000 randomly chosen points from both posteriors. Not only will a K amplitude give further weight to $\mathcal{P}1$, but it can also be seen that the $\mathcal{P}1$ region ($\gtrsim 7^\circ$) has K dependence, implying RV measurements will better constrain mutual inclination there. }
\label{fig:komega}
\end{figure}

Although we are confident that this system has a large mutual inclination, a small number of radial velocity (RV) data points could help further constrain the system's parameters. RV measurements may be able to determine which star the planets are truly around and thus refine our fit significantly. Additionally, the RV curves are vastly different in shape between the $\mathcal{P}1$ and $\mathcal{P}2$ models due to the different eccentricities in the models. If the RV curve is observed to be saw-toothed, it would also give additional constraints on $e$ and $\omega$ which are not well-measured in the photometry. Further, because the RV $K$ amplitude is dependent on eccentricity (as well as several other factors),
\begin{equation}
K = \bigg( \frac{2 \pi G}{P} (m_1+m_2) \bigg)^{1/3} \frac{m_2}{m_1}  \frac{\sin[i]}{\sqrt{1-e^2}},
\end{equation} 
the overall amplitude of the RV signal will be drastically different in the nearly coplanar case $\mathcal{P}2$ compared to $\mathcal{P}1$, allowing for additional confirmation (Fig.~\ref{fig:komega}). In addition, the K amplitude alone will help constrain the value of the mutual inclination in the highly mutually inclined case ($\mathcal{P}1$) since the K amplitude varies as a function of mutual inclination. A foreseeable challenge for RV measurements is that the two stars are only 1" apart, roughly the seeing limit for ground based observations.

\subsection{Non-transiting Planets}

So far our discussion has included only the two planets observed in transit. The transit timing variations of the two observed planets can, in principle, put constraints on the orbits of non-transiting planets. Since the observed planets have moderate eccentricities and mutual inclinations, we must consider that any unobserved planet also may also have a substantial eccentricity and mutual inclination (which may be the cause of it not transiting). While constraints on non-transiting planets in systems where circular, coplanar orbits are assumed can be quite tight \citep{2005MNRAS.359..567A,2005MNRAS.364L..96S,2007MNRAS.374..941A}, considering eccentricity to first or higher orders vastly complicates this process \citep{2015arXiv150901623A}. The addition of mutual inclinations will add further allowable TTV frequencies and amplitudes for unseen planets at a given period, and thus decomposing observed signals into the sums of transiting and hypothetical non-transiting planets to set upper limits on unseen planets of a given mass as a function of period becomes untenable. 

Since the two planets completely explain the TTVs (the residuals are consistent with no signal), we do not appeal to the existence of more planets. Additionally, more planets, particularly in a system of moderately high eccentricities and mutual inclinations, increases the chance that the system would be unstable.

The two known planets in Kepler-108 should both produce observable K amplitudes (Planet b: $\gtrsim 10 m/s$, Planet b: $\gtrsim 3 m/s$), and we expect that other Jovian-mass planets in the system with periods shorter than the outermost transiting planet ($P\approx190.3$ d) may also be detectable through RV measurements. Since we speculate that this system experienced a planet-planet scattering event, it is likely that any other planets in the system may not be coplanar with the observed ones and thus only detectable through RVs, not transits. Small ($\lesssim 0.1 M_{Jup}$) or longer period ($\gtrsim 200$d) planets would likely not be detected by RVs.

\section{Dynamical Discussion}
\label{sec:discussion}

We have presented a photodynamic analysis of the orbital parameters of the giant planet system Kepler-108. Planetary systems formed in disks are likely to be coplanar and nearly circular. However, the planets in Kepler-108 are shown to have a high mutually inclination ($\Delta \Omega \gtrsim15^\circ$) and eccentricity ($e_c \gtrsim 0.1$), not what one would expect from a purely disk formation origin. Instead this system shows signs of a more violent, chaotic past as is predicted by theories of secular chaos and the formation of hot Jupiters, establishing an observational link between theoretical stages of planetary system evolution.

The presence of an additional companion star increases the richness of the dynamics of Kepler-108. Kozai-Lidov cycles from a distant companion have been suggested as a means of exciting eccentricities of planets, which may lead to strong planet-planet interactions including scattering and ejection \citep{2007MNRAS.377L...1M}. The timescale for Kozai-Lidov cycles is
\begin{equation}
\tau = \frac{2}{3 \pi} \frac{P_{\star}^2}{P_{pl}} \frac{M_1 + M_2 + M_{pl}}{M_2} (1-e_{\star}^2)^{3/2}
\end{equation}
\citep{1998MNRAS.300..292K,2007ApJ...669.1298F}, where $1$ refers to the central star, $2$ the companion star, $pl$ the planet of interest, and $P_\star$ refers to the binary star period. We do not know the period or eccentricity of the outer star, only its sky-projected distance, which is approximately 327 AU \citep{2015ApJ...813..130W}. The true distance is likely larger because this measurement ignores the separation of the stars along the axis in the direction of the observer. RV measurements could track the change in velocity as a function of time (i.e. $a_z$, where $a$ is the acceleration and the subscript $z$ represents the direction along the line of sight), which would allow an estimate of $r_z$, since $M$ and $r_\perp$ are known, where $\perp$ denotes the sky-plane direction, by solving the following for $r_z$:
\begin{equation}
\dot{v}_z = \frac{G M }{(r_\perp^2+r_z^2)^{3/2}} r_z
\end{equation}
However, even knowing the true separation of the stars would not reveal the period of the companion star because the star's orbit may not be circular. Rather, the star may be near pericenter of much larger semi-major axis orbit or near apocenter of a much shorter, highly-eccentric orbit.  Still, we desire to understand whether or not Kozai cycles from interactions with this companion star could be influencing the dynamics of Kepler-108 system. 

\begin{figure}
\centerline{
\includegraphics[scale=1.0]{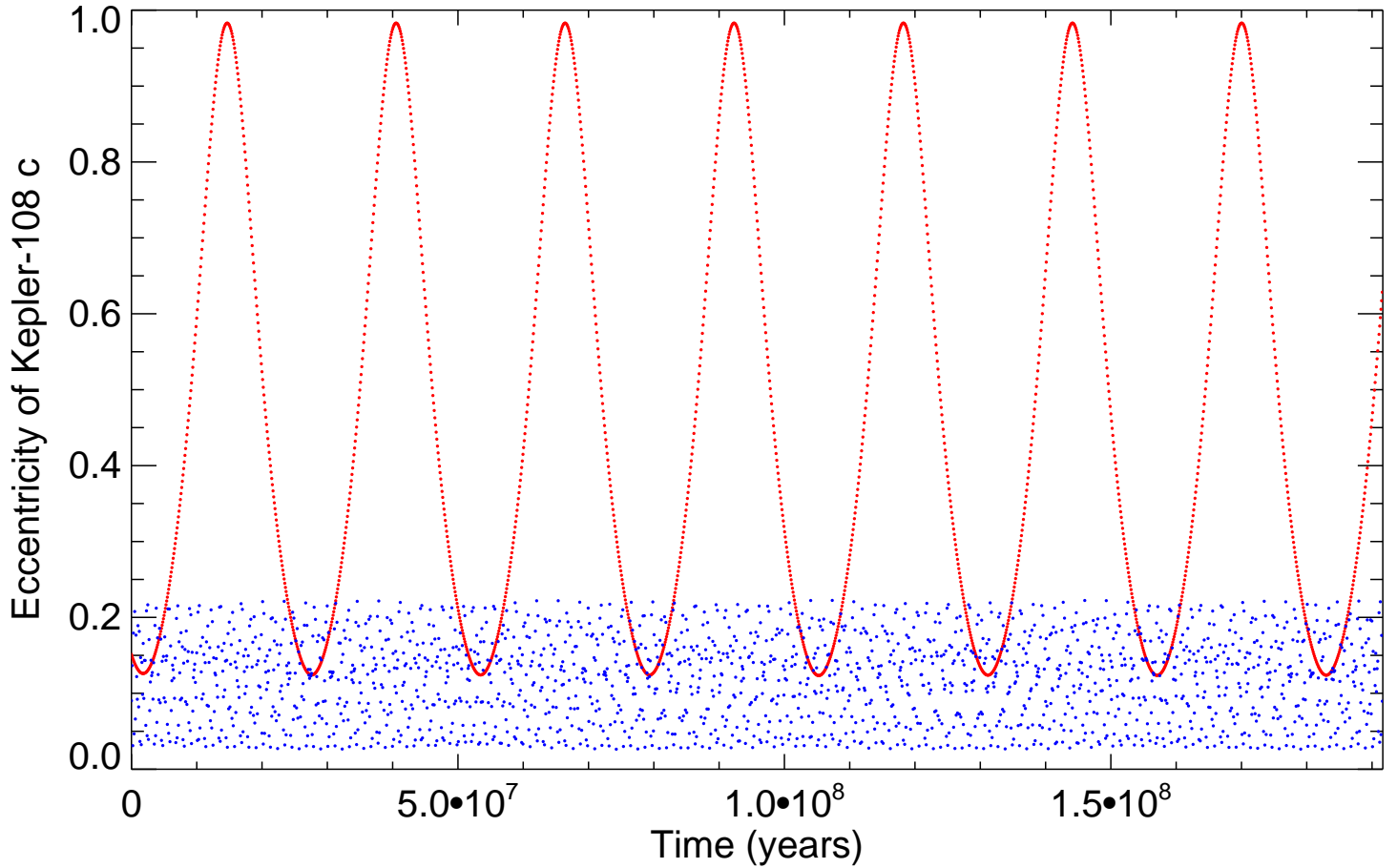}
}
\caption{
The eccentricity of Kepler-108 c as found by numerical simulation in the presence of a perturbing 1.377 $M_\odot$ star in a circular 327.5 AU orbit and $i=10^\circ$. The blue points represent the true 2-planet Kepler-108 system. The planet-planet interactions suppress the Kozai-Lidov oscillations and keep Kepler-108 c's eccentricities at moderate values. Kozai-Lidov oscillations driving Kepler-108 c are clearly present in a simulation with all other parameters identical, but not including the interior planet Kepler-108 b. These points are represented in red and show the very large and potentially destabilizing eccentricity swings that would result in Kepler-108 without taking the strong planet-planet interactions into account. 
 }
\label{fig:kozai}
\end{figure}

If we assume the binary orbit is nearly circular and has a semi-major axis approximately equal to the sky-projected distance (327 AU), we derive $P_{\star} \sim 3900$ year and $\tau \sim 10$ Myr. This means that if the inclination of the companion star to the Kepler-108 c is large, it could potentially drive Kozai-Lidov oscillations and cause strong planet-planet interactions on this timescale. It is also entirely possible that the Kozai-Lidov timescale is longer than the age of the system (in large part because the timescale depends on the extremely uncertain $P_{out}$ to the second power), or that the companion star is on a nearly coplanar orbit with the planets, in which case the Kozai-Lidov mechanism does not apply. Additionally, since the planet-planet precession interaction timescale is relatively short $P_{prec} << \tau$, this can dominate the dynamics and prevent Kozai-Lidov cycles from occurring. To test this, we ran several realizations of the system by integrating forward in time the best fit solution, with the additional companion star on a circular orbit at 327 AU, using the \texttt{MERCURY} \citep{2012ascl.soft01008C} integrator. We run the simulations for 200 Myr, many times the expected Kozai-Lidov timescale of the system. The inclination of the companion star is varied by 10 degree intervals from 0 to 180 degrees. To ensure that the Kozai-Lidov mechanism works as expected in a 3-body system, we also run the same set of simulations without the inner planet (Kepler-108 b). We find that the planet-planet interactions in the 2-planet systems dominate and do not allow Kozai-Lidov eccentricity cycles to occur (see, e.g., Fig.~\ref{fig:kozai}). All 2-planet systems tested remained stable for $2\times10^8$ years. 

It is plausible that an additional planet at a much greater orbital period than of the observed 2 planets could have been subject to Kozai-Lidov oscillations shortly after dissipation of the natal disk, reached a high eccentricity, and caused a planet-planet scattering event. This could result in the large mutual inclination of the two observed planets. However, the parameter space for unobserved, possibly-ejected, long-period planets is very large and we do not complete any numerical analysis of this scenario.

The rough similarity between the observed nodal precession timescales and the possible period of the binary star presents another intriguing possibility of the origin of this system: excitation of the planets' mutual inclination through a Laplace-Lagrange evection resonance \citep{2015Natur.524..439T}. Studying the Kepler-108 system in this context may require additional data, particularly about the nature of the stellar binary's orbit, and theory, so we leave it to future work. 


In summary, we have shown that Kepler-108 is a mutually inclined giant planet system with a well-measured precession rate through TTV and TDV analysis. 

\acknowledgements

We thank Philip Lucas for assistance in understanding the UKIRT data. This material is based upon work supported by NASA under Grant Nos. NNX14AB87G issued through the \Kepler Participating Scientist Program. D.C.F received support from the Alfred P. Sloan Foundation. Computer simulations were run using the ``Midway" cluster at University of Chicago Research Computing Center. Much of the data presented in this paper were obtained from the Mikulski Archive for Space Telescopes (MAST). STScI is operated by the Association of Universities for Research in Astronomy, Inc., under NASA contract NAS5-26555. Support for MAST for non-HST data is provided by the NASA Office of Space Science via grant NNX13AC07G and by other grants and contracts. 
The United Kingdom Infrared Telescope (UKIRT) is supported by NASA and operated under an agreement among the University of Hawaii, the University of Arizona, and Lockheed Martin Advanced Technology Center; operations are enabled through the cooperation of the Joint Astronomy Centre of the Science and Technology Facilities Council of the U.K. When the data reported here were acquired, UKIRT was operated by the Joint Astronomy Centre on behalf of the Science and Technology Facilities Council of the U.K. 
This work makes use of observations from the Las Cumbres Observatory Global Telescope Network, the Kepler Community Follow-up Observing Program (CFOP), and NASA's Astrophysics Data System (ADS).

\bibliographystyle{apj}
\bibliography{references}

\begin{table}
\centering
\begin{tabular} {l | c |  c}
\multicolumn{3}{c}{\bf Kepler-108 Stellar Properties} \\
\hline
  	& Kepler-108 A	& Kepler-108 B (Planet Host) \\
\hline
	& \multicolumn{2}{c}{ Asteroseismology$^a$}\\
\hline
$M_\star (M_\odot)$ 		&$1.377 \pm 0.089$	& -	\\
$R_\star (R_\odot)$ 		& $ 2.192 \pm 0.121$ & -	\\
\hline
	& \multicolumn{2}{c}{ Photometry}\\
\hline
$i_{mag}$$^{bd}$ & 	$12.90 \pm 0.22$	& $13.77 \pm 0.22$	\\
$J_{mag}$$^{cd}$ & 	$12.087 \pm 0.15$ 	& $12.287 \pm 0.15$ 	\\
$K_{mag}$$^{cd}$ & 	$11.640 \pm 0.15$	& $11.840 \pm 0.15$ 	\\
$M_\star (M_\odot)$$^e$ 		& 	$1.26_{-0.23}^{+0.33}$	& $0.96_{-0.16}^{+0.29}$	\\
$R_\star (R_\odot)$$^e$		& 	$1.45_{-0.41}^{+0.73}$	& $0.97_{-0.21}^{+0.56}$	\\
\end{tabular}
\caption{ 
$^a$\cite{2013ApJ...767..127H} $^b$\cite{2014ApJ...791...35L} $^c$\cite{2015ApJ...813..130W} $^d$https://cfop.ipac.caltech.edu $^e$\cite{2015ascl.soft03010M} 
}

\label{table:star}

\end{table}

\begin{table}
\centering
\scalebox{0.8}{
\begin{tabular} {l || c c c | c c c || }
\multicolumn{7}{c}{\bf Kepler-108 Posteriors} \\
\cline{2-7}
& \multicolumn{6}{ c || }{ \bf Host: Kepler-108B }  \\ 
\hline
& \multicolumn{3}{ c | }{ \bf $\mathcal{P}1$ - Mutually Inclined } & \multicolumn{3}{ c || }{ \bf $\mathcal{P}2$ - Nearly Coplanar } \\ 
\cline{2-7}
 &  Median   & 68.3\% (1-$\sigma$)  & 95.4\% (2-$\sigma$) &  Median   & 68.3\% (1-$\sigma$) & 95.4\% (2-$\sigma$)   \\
Parameter Name (Unit) &   & Uncertainties & Uncertainties &    & Uncertainties &  Uncertainties\\
\hline
\emph{Stellar Parameters:} &&&&&&  \\	
$R_\star (R_\odot)$ 		& $1.62 $ & $^{+0.16}_{-0.17}$ & $^{+0.55}_{-0.31}$			& $1.911 $ & $^{+0.082}_{-0.079}$ & $^{+0.16}_{-0.23}$ 		\\
$M_\star (M_\odot)$ 		& $0.96 $ & $^a$ &										& $0.96 $ &$^a$ & 									\\
$c_1$ 				& $0.548 $ & $^{+0.070}_{-0.058} $ & $^{+0.17}_{-0.12}$		& $0.564 $ & $^{+0.086}_{-0.079}$ & $^{+0.18}_{-0.16}$ 	 	\\
$c_2$ 				& $-0.055 $ & $^{+0.097}_{-0.098}$ & $^{+0.20}_{-0.20}$		& $-0.02 $ & $^{+0.10}_{-0.10}$  & $^{+0.22}_{-0.22}$  	\\
$dilute$ 				& $0.641$ & $^{+0.083}_{-0.19}$	& $^{+0.13}_{-0.51}$		& $0.20$ & $^{+0.20}_{-0.14}$ & $^{+0.43}_{-0.20}$				\\
\emph{Kepler-108 b Parameters:} &&&&&& \\
P (d) 					& $ 49.18351 $ & $^{+0.00034}_{-0.00059} $ & $^{+0.0013}_{-0.0012}$  	& $ 49.18360 $ & $^{+0.00012}_{-0.00016} $ & $^{+0.00023}_{-0.00050}$	\\ 
T$_0$ (BJD-2454900 (d)) 		& $ 665.12168$ & $^{+0.00061}_{-0.00077} $ & $^{+0.0012}_{-0.0040}$	& $ 665.1121 $ & $^{+0.0023}_{-0.0035} $ & $^{+0.0075}_{-0.010}$ 			\\  
$e \cdot \cos (\omega)$  		& $  -0.092 $ & $^{+0.046}_{-0.11} $ & $^{+0.12}_{-0.49}$				& $ -0.672 $ & $^{+0.015}_{-0.014} $ & $^{+0.044}_{-0.029}$ 				\\  
$e \cdot \sin (\omega)$ 		& $ -0.049 $ & $^{+0.033}_{-0.19} $ & $^{+0.12}_{-0.38}$ 				& $ -0.417 $ & $^{+0.029}_{-0.025} $  & $^{+0.16}_{-0.052}$ 				\\  
$e_{tot}$$^b$				& $ 0.13 $ & $^{+0.16}_{-0.068} $ & $^{+0.59}_{-0.10}$ 				& $ 0.792 $ & $^{+0.021}_{-0.025} $  & $^{+0.042}_{-0.11}$ 				\\
i ($^\circ$) 				& $ 90.59 $ & $^{+0.38}_{-0.22} $ & $^{+0.95}_{-0.37}$				& $ 91.82 $ & $^{+0.20}_{-0.27} $ & $^{+0.41}_{-1.0}$ 					\\  
$\Omega$ ($^\circ$) 			& $ 0.0 $ &  &		  										& $0.0$ & & 													\\  
M (M$_{jup}$) 				& $ 0.308 $ & $^{+0.16}_{-0.069} $ & $^{+0.65}_{-0.13}$ 				& $ 1.29 $ & $^{+0.33}_{-0.30} $ & $^{+0.78}_{-1.4}$ 					\\  
$\rm{R}/\rm{R}_\star$ 		& $ 0.0620 $ & $^{+0.0083}_{-0.011} $ & $^{+0.015}_{-0.020}$ 			& $ 0.0429 $ & $^{+0.0063}_{-0.0031} $  & $^{+0.019}_{-0.0042}$ 			\\  
\emph{Kepler-108 c Parameters:} &&&&&&  \\
P (d) 					& $ 190.339 $ & $^{+0.013}_{-0.0098} $ & $^{+0.048}_{-0.021}$ 		& $ 190.545 $ & $^{+0.098}_{-0.092} $ & $^{+0.21}_{-0.25}$ 				\\  
T$_0$ (BJD-2454900 (d)) 		& $ 816.657 $ & $^{+0.013}_{-0.011} $  & $^{+0.054}_{-0.026}$  		& $ 816.842 $ & $^{+0.095}_{-0.089} $ & $^{+0.20}_{-0.23}$ 				\\  
$e \cdot \cos (\omega)$  		& $ 0.051 $ & $^{+0.026}_{-0.065} $ & $^{+0.045}_{-0.14}$  			& $ -0.1223 $ & $^{+0.0041}_{-0.0038} $ & $^{+0.011}_{-0.0073}$ 			\\  
$e \cdot \sin (\omega)$  		& $ -0.125 $ & $^{+0.033}_{-0.025} $ & $^{+0.22}_{-0.06}$ 			& $ -0.2276 $ & $^{+0.0073}_{-0.0063} $ & $^{+0.067}_{-0.013}$ 			\\  
$e_{tot}$$^b$				& $ 0.14 $ & $^{+0.025}_{-0.027} $ & $^{+0.076}_{-0.083}$ 			& $ 0.2585 $ & $^{+0.0057}_{-0.0066} $  & $^{+0.011}_{-0.061}$ 				\\
i ($^\circ$) 				& $ 90.416 $ & $^{+0.082}_{-0.10} $ & $^{+0.16}_{-0.11}$				& $ 90.547 $ & $^{+0.034}_{-0.041} $ & $^{+0.067}_{-0.095}$ 				\\  
$\Omega$ ($^\circ$) 			& $ 30 $ & $^{+13}_{-9} $ & $^{+30}_{-18}$ 						& $0.0$ & & 													\\  
M (M$_{jup}$) 				& $ 0.17 $ & $^{+0.10}_{-0.094} $  & $^{+0.22}_{-0.14}$  				& $ 0.0174 $ & $^{+0.0052}_{-0.0043} $ & $^{+0.026}_{-0.0084}$ 			\\  
$\rm{R}/\rm{R}_\star$ 	 	& $ 0.0544 $ & $^{+0.0071}_{-0.0099} $ & $^{+0.013}_{-0.018}$ 		& $ 0.0388 $ & $^{+0.0058}_{-0.0028} $ & $^{+0.039}_{-0.0039}$ 			\\

\end{tabular}
}
\caption{ 
 $^a$ Note that the stellar mass is held fixed in these simulations so the values and uncertainties on the planets' masses may easily be scaled with future measurements of the stellar mass. 
 $^b$ $e_{tot}$ is not actually a fitted parameter, rather it is derived from $e \cdot \cos (\omega)$ and $e \cdot \sin (\omega)$.
}

\label{table:allparams}

\end{table}

\begin{table}
\hspace*{-2.0cm}
\scalebox{0.9}{
\setlength{\tabcolsep}{2pt}
\centering
\tiny
\begin{tabular} {c c c c c c c c c}
\multicolumn{9}{c}{Kepler-108B Best-Fit Solutions} \\
\hline
Planet & Period (d) & $T_0$ (BJD-2454900) & $e$ & $i$ ($^\circ$) & $\Omega$ ($^\circ$) & $\omega$ ($^\circ$) & Mass ($M_{Jup}$) & Radius ($R_p/R_\star$) \\
\hline
b &   49.183151935389887   &   665.121856657809190   &   0.090102588624939   &   90.472025946068158  &    0.0    &   -154.107928542839574  &  0.293493193888467  &     0.063802124701062 \\
c  &   190.338447063836668   &  816.659623505046625  &    0.151238887074790  &    90.409059655063757   &   30.266668634547806  &    -59.803981478553702    & 0.253483032889065  &     0.056568977896875 \\
\emph{Stellar Parameters:} & $M_\star$ ($M_\odot$): 0.96  & $R_\star$($R_\odot$): & 1.613083860753101  & $c_1$: 0.503653056308355 & $c_2$: 0.003501180871091 & $dilute$: & 0.666561048263810 & \\
\hline
b &   49.183612083993282   &   665.112150617523753   &   0.794341270764740    &  91.810317600437372    &  0.0    &   -147.586803924091697   & 1.365714849510034   &    0.042928495826805 \\
c  &   190.542420419310957   &  816.838109271527060   &   0.259066554493369     & 90.538697678534362   &   0.0  &     -117.928142257279774   & 0.017969797702308   &    0.038697748216860\\
\emph{Stellar Parameters:} & $M_\star$ ($M_\odot$): 0.96 & $R_\star$($R_\odot$): & 1.890968346442407 & $c_1$: 0.556644949922503 & $c_2$: -0.006100068227655 & $dilute$: & 0.205417387715994 & \\
\hline
\end{tabular}
 }
\footnotesize
\caption{
 }

\label{table:bestfit}
\end{table}

\begin{table}
\centering
\begin{tabular} {l c c}
\multicolumn{3}{c}{\bf Kepler-108 Observational Likelihood} \\
\hline
  	& Fraction of Viewing Angles	& Fraction with Planets and a \\
	& With Planets Observed		& Measurable Duration Drift  \\
\hline
Two Planets 		& $0.0005(2) $		& $0.0003(2)     $	\\
Single Planet 		& $0.0423(21) $	& $0.00134(12) $	\\
None Visible		& $0.9572(98) $	& n/a			\\
\end{tabular}
\caption{ 
}

\label{table:obsfrac}

\end{table}

\begin{table}
\centering
\tiny
\begin{tabular} {l | c | c | c | c}
\multicolumn{5}{c}{\bf Kepler-108 Transit Times} \\
\hline
& \multicolumn{2}{c|}{Kepler-108 b} & \multicolumn{2}{c}{Kepler-108 c}  \\
\hline
n	&  Time (d)		& Uncertainty (d) &  Time (d)		& Uncertainty (d)  \\
\hline
{\bf         -12 }&{\bf       74.908501 }&{\bf   0.00086			 }&{\bf }&{\bf }\\
{\bf         -11 }&{\bf       124.09398 }&{\bf   0.00075 }&{\bf }&{\bf }\\
{\bf         -10 }&{\bf       173.27824 }&{\bf   0.00081 }&{\bf }&{\bf }\\
{\bf          -9 }&{\bf       222.46238 }&{\bf   0.00068 }&{\bf }&{\bf }\\
{\bf          -8 }&{\bf       271.64709 }&{\bf   0.00054 }&{\bf }&{\bf }\\
{\bf          -7 }&{\bf       320.83224 }&{\bf   0.00069 }&{\bf }&{\bf }\\
{\bf          -6 }&{\bf       370.01746 }&{\bf    0.0011 }&{\bf }&{\bf }\\
{\bf          -5 }&{\bf       419.20065 }&{\bf   0.00067 }&{\bf }&{\bf }\\
{\bf          -4 }&{\bf       468.38466 }&{\bf   0.00057 }&{\bf }&{\bf }\\
{\bf          -3 }&{\bf       517.56880 }&{\bf   0.00089	}&{\bf	245.68203 }&{\bf    0.0025 }\\
{\bf          -2 }&{\bf       566.75487 }&{\bf    0.0014 }&{\bf    435.99307 }&{\bf    0.0018 }\\
{\bf          -1 }&{\bf       615.93712 }&{\bf   0.00077 }&{\bf    626.31352 }&{\bf    0.0017 }\\
{\bf           0 }&{\bf       665.12038 }&{\bf   0.00052 }&{\bf    816.64099 }&{\bf    0.0021 }\\
{\bf           1 }&{\bf       714.30337 }&{\bf    0.0010 }&{\bf    1006.9670 }&{\bf    0.0023 }\\
{\bf           2 }&{\bf       763.48960 }&{\bf    0.0011 }&{\bf    1197.2845 }&{\bf    0.0018 }\\
{\bf           3 }&{\bf       812.67169 }&{\bf   0.00072 }&{\bf    1387.5931 }&{\bf    0.0024  }\\
{\bf           4 }&{\bf       861.85489 }&{\bf   0.00047 }&    1577.8978 &    0.0054 \\
{\bf           5 }&{\bf       911.03735 }&{\bf    0.0010 }&    1768.2046 &    0.0088 \\
{\bf           6 }&{\bf       960.22307 }&{\bf   0.00066 }&    1958.5187 &     0.010 \\
{\bf           7 }&{\bf       1009.4058 }&{\bf   0.00064 }&    2148.8422 &    0.0096 \\
{\bf           8 }&{\bf       1058.5897 }&{\bf   0.00050 }&    2339.1702 &    0.0085 \\
{\bf           9 }&{\bf       1107.7724 }&{\bf   0.00078 }&    2529.4937 &    0.0082 \\
{\bf          10 }&{\bf       1156.9575 }&{\bf   0.00062 }&    2719.8078 &    0.0092 \\
{\bf          11 }&{\bf       1206.1412 }&{\bf   0.00053 }&    2910.1144 &     0.011 \\
{\bf          12 }&{\bf       1255.3259 }&{\bf   0.00052 }&    3100.4192 &     0.015 \\
{\bf          13 }&{\bf       1304.5091 }&{\bf   0.00057 }&    3290.7281 &     0.018 \\
{\bf          14 }&{\bf       1353.6939 }&{\bf   0.00071 }&    3481.0456 &     0.019 \\
{\bf          15 }&{\bf       1402.8784 }&{\bf   0.00058 }&    3671.3715 &     0.018 \\
{\bf          16 }&{\bf       1452.0637 }&{\bf   0.00073 }&    3861.6989 &     0.017 \\
{\bf          17 }&{\bf       1501.2474 }&{\bf   0.00078 }&    4052.0192 &     0.017 \\
          18 &       1550.4319 &   0.00095 &    4242.3303 &     0.018 \\
          19 &       1599.6168 &   0.00097 &    4432.6356 &     0.022 \\
          20 &       1648.8022 &    0.0011 &    4622.9412 &     0.025 \\
          21 &       1697.9868 &    0.0014 &    4813.2527 &     0.028 \\
          22 &       1747.1706 &    0.0013 &    5003.5736 &     0.028 \\
          23 &       1796.3551 &    0.0014 &    5193.9010 &     0.026 \\
          24 &       1845.5400 &    0.0015 & & \\
          25 &       1894.7255 &    0.0020 & & \\
          26 &       1943.9084 &    0.0016 & & \\
          27 &       1993.0921 &    0.0015 & & \\
          28 &       2042.2758 &    0.0017 & & \\
          29 &       2091.4620 &    0.0020 & & \\
          30 &       2140.6441 &    0.0015 & & \\
          31 &       2189.8273 &    0.0014 & & \\
          32 &       2239.0100 &    0.0016 & & \\
          33 &       2288.1961 &    0.0015 & & \\
          34 &       2337.3784 &    0.0013 & & \\
          35 &       2386.5617 &    0.0013 & & \\
          36 &       2435.7442 &    0.0014 & & \\
          37 &       2484.9297 &    0.0012 & & \\
          38 &       2534.1128 &    0.0011 & & \\
          39 &       2583.2970 &    0.0012 & & \\

\end{tabular}
\quad
\begin{tabular} {l | c| c}
\multicolumn{3}{c}{\bf Kepler-108 Transit Times} \\
\hline
& \multicolumn{2}{c}{Kepler-108 b}  \\
\hline
n	&  Time (d)$^a$$^b$		& Uncertainty (d)  \\
\hline
          40 &       2632.4798 &    0.0012  \\
          41 &       2681.6648 &    0.0013  \\
          42 &       2730.8488 &    0.0012  \\
          43 &       2780.0338 &    0.0014  \\
          44 &       2829.2171 &    0.0014  \\
          45 &       2878.4018 &    0.0016  \\
          46 &       2927.5865 &    0.0017  \\
          47 &       2976.7719 &    0.0019  \\
          48 &       3025.9559 &    0.0020  \\
          49 &       3075.1402 &    0.0021  \\
          50 &       3124.3250 &    0.0023  \\
          51 &       3173.5103 &    0.0024  \\
          52 &       3222.6952 &    0.0027  \\
          53 &       3271.8787 &    0.0026  \\
       54 &       3321.0630 &    0.0026  \\
          55 &       3370.2475 &    0.0028  \\
          56 &       3419.4333 &    0.0031  \\
          57 &       3468.6158 &    0.0028  \\
          58 &       3517.7993 &    0.0027  \\
          59 &       3566.9826 &    0.0028  \\
          60 &       3616.1689 &    0.0029  \\
          61 &       3665.3509 &    0.0026  \\
          62 &       3714.5341 &    0.0026  \\
          63 &       3763.7166 &    0.0027  \\
          64 &       3812.9025 &    0.0025  \\
          65 &       3862.0851 &    0.0024  \\
          66 &       3911.2687 &    0.0024  \\
          67 &       3960.4513 &    0.0024  \\
          68 &       4009.6366 &    0.0024  \\
          69 &       4058.8200 &    0.0023  \\
          70 &       4108.0044 &    0.0024  \\
          71 &       4157.1874 &    0.0024  \\
          72 &       4206.3723 &    0.0025  \\
          73 &       4255.5566 &    0.0026  \\
          74 &       4304.7417 &    0.0028  \\
          75 &       4353.9253 &    0.0028  \\
          76 &       4403.1099 &    0.0030  \\
          77 &       4452.2946 &    0.0031  \\
          78 &       4501.4801 &    0.0033  \\
          79 &       4550.6643 &    0.0035  \\
          80 &       4599.8484 &    0.0035  \\
          81 &       4649.0331 &    0.0036  \\
          82 &       4698.2182 &    0.0038  \\
          83 &       4747.4034 &    0.0041  \\
          84 &       4796.5866 &    0.0039  \\
          85 &       4845.7706 &    0.0039  \\
          86 &       4894.9547 &    0.0040  \\
          87 &       4944.1407 &    0.0042  \\
          88 &       4993.3230 &    0.0039  \\
          89 &       5042.5063 &    0.0039  \\
          90 &       5091.6893 &    0.0040  \\
          91 &       5140.8755 &    0.0039  \\
\end{tabular}

\caption{ 
$^a$ (BJD-2454900)
$^b$ TTVS measured over the duration of the \Kepler observing window are emboldened while future predicted TTVs are roman.
}

\label{table:ttvs}
\end{table}

\clearpage
\appendix
\section{Posteriors with Kepler-108A as the Planetary Host}
\clearpage

\begin{table}
\centering
\scalebox{0.8}{
\begin{tabular} {l || c c c | c c c || }
\multicolumn{7}{c}{\bf Kepler-108 Posteriors$^b$} \\
\cline{2-7}
& \multicolumn{6}{ c || }{ \bf Host: Kepler-108A }  \\ 
\hline
& \multicolumn{3}{ c | }{ \bf $\mathcal{P}1$  - Mutually Inclined } & \multicolumn{3}{ c || }{ \bf $\mathcal{P}2$ - Nearly Coplanar} \\ 
\cline{2-7}
 &  Median   & 68.3\% (1-$\sigma$)  & 95.4\% (2-$\sigma$) &  Median   & 68.3\% (1-$\sigma$) & 95.4\% (2-$\sigma$)   \\
Parameter Name (Unit) &   & Uncertainties & Uncertainties &    & Uncertainties &  Uncertainties\\
\hline
\emph{Stellar Parameters:} &&&&&&  \\
$R_\star (R_\odot)$ 		& $2.13 $ & $^{+0.14}_{-0.13}$ & $^{+0.27}_{-0.25}$		& $2.188 $ & $^{+0.070}_{-0.074}$ & $^{+0.13}_{-0.15}$ 		\\
$M_\star (M_\odot)$ 		& $1.377 $ & &								& $1.377 $ & & 									\\
$c_1$ 				& $0.579 $ & $^{+0.080}_{-0.073} $ & $^{+0.17}_{-0.14}$	& $0.576 $ & $^{+0.086}_{-0.083}$ & $^{+0.17}_{-0.17}$ 	 	\\
$c_2$ 				& $-0.06 $ & $^{+0.10}_{-0.10}$ & $^{+0.21}_{-0.21}$	& $-0.02 $ & $^{+0.10}_{-0.10}$  & $^{+0.22}_{-0.22}$  	\\
$dilute$ 				& $0.44$ & $^{+0.15}_{-0.21}$	& $^{+0.25}_{-0.39}$		& $0.15$ & $^{+0.14}_{-0.10}$ & $^{+0.27}_{-0.14}$				\\
\emph{Kepler-108 b Parameters:} &&&&&& \\
P (d) 					& $ 49.18358 $ & $^{+0.00021}_{-0.00049} $ & $^{+0.0018}_{-0.0012}$  	& $ 49.18360 $ & $^{+0.00012}_{-0.00014} $ & $^{+0.00023}_{-0.00031}$	\\ 
T$_0$ (BJD-2454900 (d)) 		& $ 665.12107 $ & $^{+0.00082}_{-0.0030} $ & $^{+0.0012}_{-0.0032}$	& $ 665.1119 $ & $^{+0.0020}_{-0.0034} $ & $^{+0.0034}_{-0.011}$ 						\\  
$e \cdot \cos (\omega)$  		& $ -0.078 $ & $^{+0.046}_{-0.122} $ & $^{+0.07}_{-0.44}$			& $ -0.672 $ & $^{+0.014}_{-0.014} $ & $^{+0.030}_{-0.030}$ 							\\  
$e \cdot \sin (\omega)$ 		& $ -0.017 $ & $^{+0.060}_{-0.049} $ & $^{+0.12}_{-0.29}$ 			& $ -0.416 $ & $^{+0.024}_{-0.025} $  & $^{+0.049}_{-0.054}$ 					\\  
i ($^\circ$) 				& $ 91.07 $ & $^{+0.20}_{-0.22} $ & $^{+0.44}_{-0.46}$				& $ 91.90 $ & $^{+0.17}_{-0.17} $ & $^{+0.35}_{-0.35}$ 								\\  
$\Omega$ ($^\circ$) 			& $ 0.0 $ &  &		  										& $0.0$ & & 																	\\  
M (M$_{jup}$) 				& $ 0.38 $ & $^{+0.26}_{-0.10} $ & $^{+0.58}_{-0.16}$ 				& $ 1.81 $ & $^{+0.37}_{-0.33} $ & $^{+0.79}_{-0.65}$ 						\\  
$\rm{R}/\rm{R}_\star$ 		& $ 0.0508 $ & $^{+0.0085}_{-0.0070} $ & $^{+0.017}_{-0.010}$ 		& $ 0.0417 $ & $^{+0.0037}_{-0.0022} $  & $^{+0.0082}_{-0.0031}$ 			\\  
\emph{Kepler-108 c Parameters:} &&&&&&  \\
P (d) 					& $ 190.337 $ & $^{+0.011}_{-0.012} $ & $^{+0.023}_{-0.023}$ 			& $ 190.543 $ & $^{+0.088}_{-0.077} $ & $^{+0.19}_{-0.14}$ 					\\  
T$_0$ (BJD-2454900 (d)) 		& $ 816.657 $ & $^{+0.012}_{-0.014} $  & $^{+0.024}_{-0.027}$  		& $ 816.841 $ & $^{+0.085}_{-0.075} $ & $^{+0.18}_{-0.14}$ 					\\  
$e \cdot \cos (\omega)$  		& $ 0.059 $ & $^{+0.024}_{-0.050} $ & $^{+0.04}_{-0.14}$  			& $ -0.1222 $ & $^{+0.0037}_{-0.0036} $ & $^{+0.0076}_{-0.0072}$ 			\\  
$e \cdot \sin (\omega)$  		& $ -0.127 $ & $^{+0.193}_{-0.027} $ & $^{+0.25}_{-0.05}$ 			& $ -0.2272 $ & $^{+0.0058}_{-0.0061} $ & $^{+0.012}_{-0.013}$ 			\\  
i ($^\circ$) 				& $ 90.549 $ & $^{+0.074}_{-0.059} $ & $^{+0.16}_{-0.11}$			& $ 90.559 $ & $^{+0.026}_{-0.027} $ & $^{+0.051}_{-0.057}$ 				\\  
$\Omega$ ($^\circ$) 			& $ 28 $ & $^{+17}_{-11} $ & $^{+32}_{-18}$ 						& $0.0$ & & 											\\  
M (M$_{jup}$) 				& $ 0.207 $ & $^{+0.093}_{-0.076} $  & $^{+0.19}_{-0.14}$  			& $ 0.0250 $ & $^{+0.0063}_{-0.0060} $ & $^{+0.013}_{-0.012}$ 			\\  
$\rm{R}/\rm{R}_\star$ 	 	& $ 0.0450 $ & $^{+0.0072}_{-0.0061} $ & $^{+0.014}_{-0.009}$ 		& $ 0.0377 $ & $^{+0.0034}_{-0.0020} $ & $^{+0.0075}_{-0.0028}$ 				\\

\end{tabular}
}
\caption{ 
$^b$The same as Table \ref{table:allparams}, except with Kepler-108A as the host star, which is strongly disfavored (\S2.2).
}

\label{table:allparamsa}

\end{table}

\begin{table}
\hspace*{-2.0cm}
\scalebox{0.9}{
\setlength{\tabcolsep}{2pt}
\centering
\tiny
\begin{tabular} {c c c c c c c c c}
\multicolumn{9}{c}{Kepler-108A Best-Fit Solutions$^a$} \\
\hline
Planet & Period (d) & $T_0$ (BJD-2454900) & $e$ & $i$ ($^\circ$) & $\Omega$ ($^\circ$) & $\omega$ ($^\circ$) & Mass ($M_{Jup}$) & Radius ($R_p/R_\star$) \\
\hline
b &   49.182439057302538   &   665.121878547496294   &   0.026651238202980    &  91.087413098342722   &   0.0    &   -174.646479043039761   & 0.466345873363698    &   0.049185434579569 \\
c  &   190.351915968598036   &  816.673491960046704    &  0.152298191712981   &   90.558802287243665   &   14.145926074400162    &  -65.774150694474812   &  0.305393885454686    &   0.044175297565015 \\
\emph{Stellar Parameters:} & $M_\star$ ($M_\odot$): 1.377 & $R_\star$($R_\odot$): & 2.157081977289178 & $c_1$: 0.609984212698404 & $c_2$: -0.096212290796273 & $dilute$: & 0.403482027905759 & \\
\hline
b &   49.183652758594832    &  665.111132211147947    &  0.791358522724910    &  91.995485644527832   &   0.0    &   -147.489191913324419 &   1.650197632313160   &    0.039137432379985 \\
c  &   190.500558445020715   &  816.801337403793127     & 0.257734476141228     & 90.569633511456729    &  0.0     &  -117.677617171142799   & 0.022451648337189   &    0.035250604625665\\
\emph{Stellar Parameters:} & $M_\star$ ($M_\odot$): 1.377  & $R_\star$($R_\odot$): & 2.217842431662132 & $c_1$: 0.542783990522848 & $c_2$: 0.032605960265989 & $dilute$: & 0.027027252762561 & \\
\hline
\end{tabular}
}
\footnotesize
\caption{ 
$^a$The same as Table \ref{table:bestfit}, except with Kepler-108A as the host star. The $\chi^2$ values here are $609824$ and $609848$ for the top and bottom parameters respectively. }
\label{table:bestfita}
\end{table}

\end{document}